\documentclass[twocolumn]{aastex631}

\usepackage{aas_macros}
\usepackage{amsmath}
\usepackage[multidot]{grffile}
\usepackage{graphics}
\usepackage{float}

\shorttitle{The Instantaneous Redshift Difference}
\shortauthors{Wang, Bolejko \& Lewis}

\begin{document}

\title{The Instantaneous Redshift Difference of Gravitationally Lensed Images: \\ Theory and Observational Prospects}
 
\author{Chengyi Wang}
\affiliation{Sydney Institute for Astronomy, School of Physics, A28, The University of Sydney, NSW 2006, Australia}

\author{Krzysztof Bolejko}
\affiliation{School of Natural Sciences, College of Sciences and Engineering, University of Tasmania, Private Bag 37, Hobart, TAS 7001, Australia}

\author{Geraint F. Lewis}
\affiliation{Sydney Institute for Astronomy, School of Physics, A28, The University of Sydney, NSW 2006, Australia}

\begin{abstract}
Due to the expansion of our Universe, the redshift of distant objects changes with time. Although the amplitude of this {\it redshift drift} is small, it will be measurable with a decade-long campaigns on the next generation of telescopes.
Here we present an alternative view of the redshift drift which captures the expansion of the universe in single epoch observations of the multiple images of gravitationally lensed sources.
Considering a sufficiently massive lens, with an associated time delay of order decades, simultaneous photons arriving at a detector would have been emitted decades earlier in one image compared to another, leading to an instantaneous {\it redshift difference} between the images. We also investigate the effect of peculiar velocities on the redshift difference in the observed images. 
Whilst still requiring the observational power of the next generation of telescopes and instruments, the advantage of such a single epoch detection over other redshift drift measurements is that it will be less susceptible to systematic effects that result from requiring instrument stability over decade-long campaigns.  
\end{abstract}


\section{Introduction}
Modern cosmology is built on the concept of an expanding universe and, whilst there is overwhelming evidence for expansion of the universe, direct detection of this phenomenon is challenging. 
The direct detection will require a measurement of the evolving redshift of a cosmological source. This is known as the {\em redshift drift} and is of order of
$10^{-18} s^{-1}$~\citep{1961ApJ...133..355S_Sandage,1962ApJ...136..319S_Sandage,1962ApJ...136..334M_McVittie, Loeb_1998}. 
However, such a direct detection would also provide a new cosmological probe in determining accelerating expansion~\citep{Uzan_2004, Uzan_2007},   
 constraining dark energy \citep{Zhang_2007,Lake_2007,Balbi_2007} and possible temporo-spatial variation of cosmological parameters  \citep{Molaro_2005,Gordon_2007,Geng_2018,amendola2021measuring}. In an extensive study, \citet{2008MNRAS.386.1192L_Liske} demonstrated that the redshift drift will be observable using next-generation of telescopes, such as the Extreme Large Telescope (ELT), through a 20-year campaign of monitoring absorption lines of hundreds quasars~\citep{2008MNRAS.386.1192L_Liske, Kim_2015,2022EJPh...43c5601Melia,Cooke_2019}.

Here we discuss the possibility of taking advantage of the gravitational lensing phenomenon to identify the redshift drift through single epoch observations of gravitationally lensed sources. We take advantage of gravitational lensing time delays between the observed images~\citep{1986ApJ...310..568Blandford}. 
This means that photons arriving at our detector today from different images were not emitted at the same instant \citep{1964MNRAS.128..295Refsdal,1992ApJ...385..404Press,10.1046/j.1365-8711.1999.02309.xBiggs}. This implies different scale factors and therefore different redshifts at these different emission instants. 
Consequently, we expect a {\em redshift difference}, $ \Delta z = \frac{dz}{dt} \, \Delta t_s$, between the images, where $\Delta t_s$ is the time interval at the source. Depending on the parameters of the lens (eg. velocity dispersion, size of the core, and position concerning the source), these time delays range from several days to a hundred years \citep{1989MNRAS.238...43K_Christopher,1994A&A...286..357Kormann}. Thus, instead of a 20-year-long campaign, we advocate targeting gravitationally lensed systems with long time delays to provide {\em direct} support for the expansion of the universe. \par

The structure of this {\em Paper} is as follows:
Sec. \ref{section:Gravitational lens theory} discusses key concepts of the gravitational lensing; Sec. \ref{section:redshift drift} introduces the redshift difference; Sec. \ref{section:observation} discusses the observability; and Sec. \ref{section:conclusion} concludes this {\em Paper}.\par

\begin{figure}
	\includegraphics[width=0.9\columnwidth]{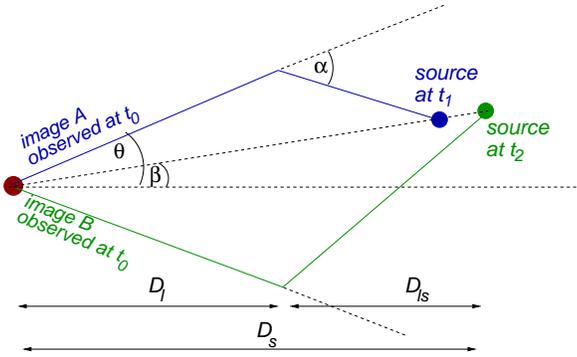}
    \caption{A schematic representation of the redshift difference. Photons propagating along different paths form multiple images in gravitational lens systems experience different time delays. Hence photons observed today in individual images were emitted at different time instances (here $t_1$ and $t_2$). Consequently, the observed images will exhibit, at the observation instant $t_0$, a different redshift.}
    \label{fig:redshift difference diagram}
\end{figure}

\section{Gravitational lens theory} \label{section:Gravitational lens theory}

Gravitational lensing, the deflection of light due to the presence of a gravitational field, can result in multiple paths connecting a source and an observer.
Different paths and also different gravitational potential along these paths, 
result in time delays. The time delay is given by \citep{1986ApJ...310..568Blandford}
\begin{equation}
    \Delta t = (1+z_l) \frac{D_l D_s}{c D_{ls}}
    \left[ \frac{1}{2}({\beta} - {\theta})^2 - \phi(\theta) \right],
    \label{Eq:time delay}
\end{equation}
where $z_l$ and $z_s$ are the redshifts of the lens and source, $D_{l}$ and $D_s$ are the corresponding angular diameter distances, whilst $D_{ls}$ is the distance between the lens and source. The angle ${\theta}$ is the position of the image and ${\beta}$ is the undeflected location of the source (cf. Fig. \ref{fig:redshift difference diagram}), and $\phi$ is the two-dimensional gravitational potential. 

The difference in time delays between two images, at the source is 
\begin{equation}
    \Delta t_{12} = \frac{D_l D_s}{c D_{ls}}\frac{1+z_l}{1+z_s} \left[\tau({\theta_2},{\beta})-\tau({\theta_1},{\beta}) \right],
    \label{Eq:time delay difference at source}
\end{equation}
where ${\theta_1}$ and ${\theta_2}$ refer to two individual images, and 
$\tau({\beta}, {\theta}) \equiv \frac{1}{2}({\beta} - {\theta})^2 - \phi({\theta})$.
For the purposes of this study we adopt the non-singular ellipsoidal lens model from~\cite{1989MNRAS.238...43K_Christopher} to represent the gravitational potential
\begin{eqnarray}
    \phi(\theta)= b  \sqrt{ s^2+(1-\varepsilon){\theta'}_x^2+(1+\epsilon){\theta'}_y^2 },
\label{Eq:Kochanek}
\end{eqnarray}
where $\theta'=(\theta'_x,\theta'_y)$ are coordinates centered at the lens and related to the sky coordinates $\theta=(\theta_x, \theta_y)$ by a translation $(\theta_{0x}, \theta_{0y})$ and a contour-clock rotation $\theta_0$ in the lens plane. The quantity $s$ is the smoothing scale, $\varepsilon$ is the eccentricity component of the lens, and the $\sigma_v$ is the velocity dispersion of the lens. The value outside the root $b=4\pi \frac{D_{ls}}{D_{s}}\left( \frac{\sigma_{v}}{c} \right)^2$, where $D_{ls}$, $D_{s}$ are angular diameter distance, $\sigma_v$ is velocity dispersion and $c$ is speed of light. Throughout we assume a cosmological model with $H_o = 67.4$ ${\rm km} / {\rm s} / {\rm Mpc}$,  $\Omega_M = 0.315$, and $\Omega_\Lambda=0.685$  \citep{refId0_Planck}.\par

\begin{figure}
	\includegraphics[width=0.9\columnwidth]{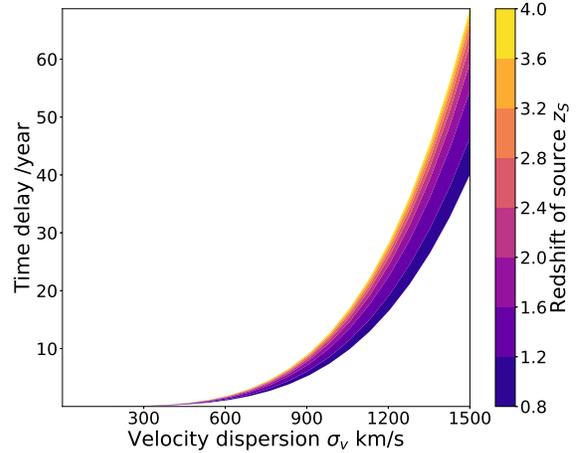}
    \caption{The change of time delay at source between two images as a function of velocity dispersion $\sigma_{v}$ and redshift of the source. The presented time delay corresponds to a maximal difference, i.e. it is a difference between the time delay of an image with the longest and the shortest time delay. The lens is located at $z_l = 0.5$.}
    \label{fig:time delay}
\end{figure}

\section{Redshift difference and gravitational lensing} \label{section:redshift drift}

\subsection{The redshift difference in a stationary system}
In a flat universe with a cosmological constant the evolution of  the scale factor has an analytical solution of the form \citep{2005MNRAS.362..213B} 
\begin{equation}
         a(t) = \bigg( \frac{\Omega_M}{\Omega_{\Lambda}} \bigg)^{1/3} \bigg[  \sinh \bigg(  \frac{3}{2}\sqrt{\Omega_{\Lambda}}H_0 t \bigg) \bigg]^{2/3} \, . \label{Eq:scalar factor}
\end{equation}
\par As noted previously, the goal here is to identify the intrinsic difference in emission times for photons arriving simultaneously at the observer (cf. Fig. \ref{fig:redshift difference diagram}). Assuming that the time difference at the source for a pair of images observed today is  $\Delta t_{12}$,  the expected redshift difference  images is
\begin{eqnarray}
        \Delta z_{12} &&= z_{2} - z_{1}
        = \frac{1}{a(t_1 + \Delta t_{12})} - \frac{1}{a(t_1)}
        \nonumber \\
         &&   \approx   - (1+z_l) H(z_s) \frac{D_l D_s}{c D_{ls}} \left[ \tau({\theta_2},{\beta})-\tau({\theta_1},{\beta}) \right],
    \label{Eq:redshift difference}
\end{eqnarray}
where $H(z_s)$ is the Hubble constant at redshift $z_s$, other parameters are same in Eq. \ref{Eq:time delay difference at source}.\par
\begin{figure}
    \centering
    \includegraphics[width=0.9\columnwidth]{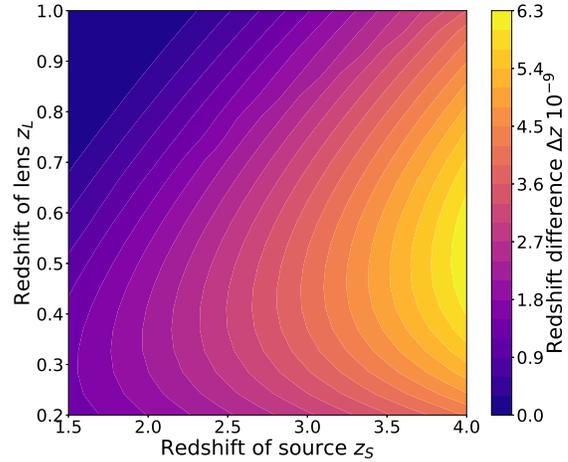}
    \caption{The redshift difference as a function of $z_l$ and $z_s$. The parameters of the system:  $s=11.9$, $\varepsilon=0.18$, $\sigma_{v} = 1000$ km/s and a source located at $\beta = (0",0")$.}
    \label{fig:redshift difference}
\end{figure}

  
In deriving the above expression, with have assumed that the differing scale factors at emission is the dominant influence on the observed redshift of an image. Clearly, there will be higher order effects in the journey of the photons, with them encountering the lensing mass at different times which will influence other aspects of the gravitational lensing configuration, such as the angular diameters distances.  However, these are the second-order effects: as discussed below and as presented in Fig. \ref{fig:redshift difference}, the redshift difference turns out to be very small. Even for a large cluster it is $\Delta z \sim 10^{-8}$, and substantially smaller for a typical galaxy. Consequently $\Delta z \Delta \tau \ll z \Delta \tau   \Rightarrow \ (1+z + \Delta z) \Delta \tau \approx (1+z ) \Delta \tau $. 
Similarly, changes in the distance $\Delta D \Delta \tau$ result in the second-order corrections. Consequently, we treat the product $H_s D_l D_s/D_{ls}$ to be the same for both images. \par

In addition, here we neglect the impact of the redshift difference due to the change of the velocity and the peculiar velocity of the lens itself. The time difference, either at the lens or source, is less than 100 years. Considering the result from \cite{2008PhLB..660...81Amendola,2021JCAP...09..018Dam}, the peculiar acceleration is less than $1\  \rm{cm/s}$ per decade for a cosmic object, which means redshift difference around $\sim 10^{-10}$. 

\begin{figure}
    \centering
    \includegraphics[scale=0.5]{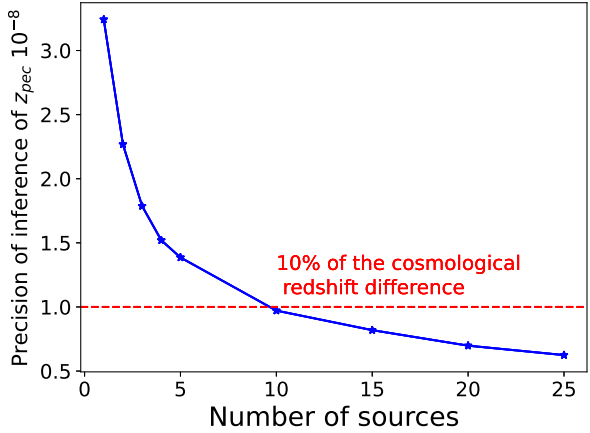}
    \caption{Errors of redshift due to peculiar velocity changes with the number of sources in a cluster lensing system moving with the velocity $v_l=1000\ \rm{km/s}$. The velocity dispersion of the lens $\sigma_v = 1500\ \rm{km/s}$ with $s=0.1$, $\varepsilon=0.7$, and $z_l=0.5$. The presented precision corresponds to the error with which the velocity of the lens can be inferred from a system with multiple lensed sources.}
    \label{fig:error_N}
\end{figure}

A more important effect absent in Eq. \ref{Eq:redshift difference}
is the Birkinshaw--Gull effect \citep{1983Natur.302..315Birkinshaw}
Due to the motion of the lens, the gravitationally lensed image will exhibit
frequency shift proportional to the lens peculiar velocity and angular separation between the image and the lens \cite{1983Natur.302..315Birkinshaw,2003ApJ...586..731Molnar_Birkinshaw,2010MNRAS.402..650Kill}
\begin{equation}
\Delta z_{pec} = \beta \gamma  \alpha\sin\delta \cos\chi,
\label{Eq:Birkinshaw-Gull}
\end{equation}
where $\beta = v_l /c$ is the ratio of lens velocity and speed of light, $\gamma = (1 - \beta^2)^{-1/2}$, $\alpha$ is the deflection angle, $\delta$ is the angle between lens velocity and line of sight, and $\chi$ is the direction of the velocity in lens plane.
This is the dominant effect.  For example, for a massive cluster moving at 300 km/s and producing gravitationally lensed images with an angular separation of $20"$, Eq. (\ref{Eq:Birkinshaw-Gull}) yields 
$\Delta z_{pec} \sim 10^{-7}$, which is approximately one magnitude larger than the expected redshift difference inferred from Eq. (\ref{Eq:redshift difference}). However, for a massive cluster we expect to observed multiple sources. With each source having multiple images, one can use Eq. (\ref{Eq:Birkinshaw-Gull}) to infer the velocity of lens and hence remove this effect from the data. To show that it can be done, we generated a mock data that comprises of randomly positioned sources around a lens of $s=0.1$, $\varepsilon=0.7$, and $\sigma_v = 1500\ \rm{km/s}$, located at located at $z_l = 0.5$, and moving with a peculiar velocity of amplitude  $v_l=1000\ \rm{km/s}$. We then adopted the position precision of the ELT of $5\times 10^{-3}\  \rm{arcsec}$ \citep{2021Msngr.182...27M_HIRES} and implemented the MCMC methods using the code 
\texttt{emcee} \citep{AKERET201327Hammer, Foreman_Mackey_2013emcee} to infer $v_{lens}$ and its accompanying error. The results are presented in Fig. \ref{fig:error_N} show that at least 10 source will be required to infer the velocity of the lens and subtract it from the signal. 

In should be however noted that our model is based on a symmetrical potential of the form given by Eq. (\ref{Eq:Kochanek}).
Realistic clusters comprise of a large number of individual galaxies making its potential less symmetrical. In addition the individual peculiar velocities of these galaxies will also contribute the redshift difference. Thus, 
in practical applications the number of required sources sufficient to minimise the noise due to the Birkinshaw–Gull effect will most likely be larger than 10. 
As a conclusion, the effects of peculiar velocity is either negligible or can be subtracted with the sufficient precision. Thus, in following discussion, we only consider the stationary situation.

\begin{figure}
    \centering
    \includegraphics[width=0.9\columnwidth]{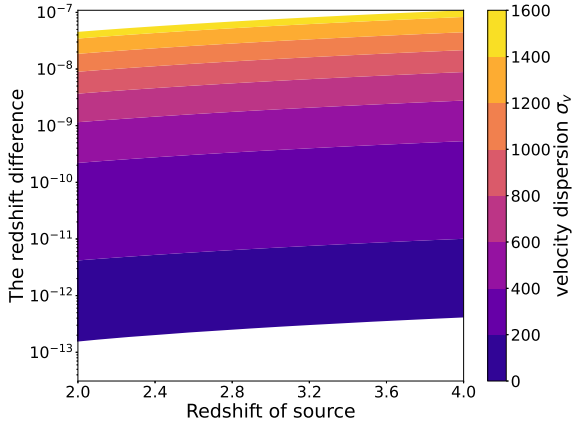}
    \caption{The redshift difference as a function of velocity dispersion. The parameters of the system:  $z_l = 0.5$, $s=0.1$, $\varepsilon=0.14$, and a source located at $\beta = (0",0")$.}
    \label{fig:redshift difference sigma}
\end{figure}

\section{Observational prospects} 
\label{section:observation} 

When addressing the question observability
we adopt a method based on \citet{2008MNRAS.386.1192L_Liske}, which was pioneered by \citet{1985Ap&SS.110..211C} and later developed by \citet{2001A&A...374..733B}. We use $\lambda_1$ and $\lambda_2$ to refer to the observed wavelength of the first and second images from which the redshift is determined. With Eqn. 16 and Fig. 12 in \citet{2008MNRAS.386.1192L_Liske}, we calculate the requirements we need to measure the redshift difference, with the main factors affecting the detectability being: (i) the amplitude of the effect, directly related to the mass of the object (hence the gravitational time delay), and (ii) the technical properties of the instrument itself. \par
 
Firstly, for a galaxy lens, recently reported by \cite{2019Bettoni}, where are $z_l = 0.56$, $z_s = 3.03$, $\sigma_{v} = 197.9$ km/s, the maximal redshift difference is of order $\Delta z = 5 \times 10^{-12}$. Using a specific emission line, such as $\lambda = 1549.48$ \AA~CIV line, the largest wavelength difference in this system is $\Delta \lambda = 7.75\times 10^{-9}$ \AA. Thus, the redshift difference induced by a typical galaxy is beyond the detectability of current or even next-generation telescopes whose accuracy is expected to be of order of $\Delta \lambda/\lambda \approx 10^{-9}$ \citep{2008MNRAS.386.1192L_Liske}.\par

However, for a galaxy cluster with velocity dispersion $\sigma_v = 1500\  \rm{km/s}$, and eccentricity parameter  $\varepsilon = 0.6$ (cf. \ref{Eq:Kochanek}), and $z_l = 0.5$ and source redshift $z_s = 4$, the expected redshift difference can be as large as 
$\Delta z = 2.67 \times 10^{-8}$. The largest wavelength difference in this system will be $\Delta \lambda = 4.14 \times 10^{-5}$ \AA \  for CIV line. This substantial difference in wavelength should be in reach of not only the next generation of optical telescopes, but also in the radio.   
 Recently, it was reported that the Five-hundred-metre Aperture Spherical radio Telescope will be, in principle, able to analyse the spectrum of observed objects at 1.4 GHz with a precision of the order of $10^{-7}$ \citep{2022PDU....3701088L}.
Although such measurements are challenging, and in addition, non-uniform matter distribution
within absorbing clouds will contribute to the uncertainty of the redshift difference, in principle this opens a new possibility to detecting the expansion of the Universe.

The main idea behind this methods is that instead of a decade-long campaign we need to target systems where the time delay is at least of order of a decade at the source. 
To maximize the redshift between gravitationally lensed images we need to maximize the interval between the emission instants of the photons that are detected today.
This is affected by two features: (i) the cosmological configuration, i.e. the redshift of the lens $z_l$ and source $z_s$, which in turns affects distances, and (ii) the characteristics of a gravitational lens. 

The impact of the cosmological configuration (i.e. redshift of lens $z_l$ and source $z_s$) for a cluster with a velocity dispersion $\sigma_v = 1000$ km/s presented in Fig. \ref{fig:redshift difference} implies that the preferable redshift of the lens is $z_l = 0.5$ and a source $z_s > 3$. 
As for the characteristics of the lens, the main parameter  influencing the amplitude of the redshift difference is the mass of the lens, encoded in the velocity dispersion of the adopted model.
Other parameters such as $\epsilon$ and $s$ play a lesser role. 
For example, fixing $z_l = 0.5$ and $z_s=4$, and changing the core radius $s$ by an order of magnitude, i.e.  from $s = 0.1"$ to 
$s=11.9"$ (cf. \cite{1995ApJ...441...58Wallington_Christopher}) results in a change of the redshift difference  from from approximately  $\Delta z = 5.31 \times 10^{-9}$ 
to approximately $\Delta z = 5.65 \times 10^{-9}$. On the other hand the change in the velocity dispersion results in more significant changes to way larger effect. 
This is presented in Fig. \ref{fig:redshift difference sigma}, where
for example the maximal redshift difference 
changes from $\Delta z = 4.7 \times 10^{-13}$ to $\Delta z = 5.6 \times 10^{-9}$ when the velocity dispersion changes from  $\sigma_{v} =100\  {\rm km/s}$  to $\sigma_{v}=1000\  {\rm km/s}$.

Thus, in order to observe the redshift difference we need to target cosmological objects with velocity dispersion larger than  $\sigma_{v}=1000\  {\rm km/s}$, which puts us in the range of massive clusters with $M > 10^{15} M_\odot$.
In order to maximise the effect of the redshift difference, the source redshift should be as large as possible, preferably $z_s \approx 4$,  and the lens redshift should preferably  be  between $z_l = 0.4$ and $z_l = 0.7$. 
With such large clusters it is expected that there will be multiple background sources being lensed. With at least 10 lensed sources we will be able to remove the effect of the peculiar velocity of the lens. This implies that the suggested in this {\em Paper} phenomenon should be observable with next generation telescopes.

\section{Summary and perspective} \label{section:conclusion}
Due to cosmic expansion, the observable properties of 
cosmological objects will change with time. The most studied effect is the redshift drift \citep{2007MNRAS.382.1623B,2008PhRvD..77b1301U}, but other properties include position drift \citep{2010PhRvD..81d3522Q,2011PhRvD..83h3503K}, flux drift \citep{2019arXiv190704495B}, the distance drift \citep{2018JCAP...03..012K,2018arXiv181110284G}, as well as a drift of gravitationally lensed images \citep{2017PhRvD..95j1301P,2022MNRAS.513.5198C}.
Here, we have discussed a new phenomenon, the {\em redshift difference} between the images of gravitationally lensed systems. This method will allow to measure the effect of the cosmological expansion in a single observation. 

 We find that the redshift difference is more sensitive to the cosmological locations of the source and the lens, as well as the lensing mass, with more massive lenses producing larger time delays and hence redshift difference. Hence, cluster lenses with time delays greater than decades make ideal targets for such a study.

It is important to emphasise that the redshift difference is a directly measurable variable of the expanding history of the universe. 
Although this will be a technically difficult observation, it has a potential to lead a measurement akin to the redshift drift.
The advantage of the redshift difference, as opposed to the classical measurement of the redshift drift,
is that it does not require a decade-long observational campaigns. By measuring the redshift difference in the gravitational lens, this long period of observation may shrink into a matter of days. 
This will eliminate problems related to keeping the instrument stable over such long periods. 
Thus, it will make the measurement of the 
redshift difference more accessible than the redshift drift.  New instruments such as FAST, SKA, and eventually ELT will thus allow us to use this new method to study the evolution of the universe at high redshift.

\begin{acknowledgments}
KB acknowledges support from the Australian Research Council through the Future Fellowship FT140101270. GFL received no funding support for this research.
\end{acknowledgments}



\end{document}